%% file: manuscript_JHEP.tex
\documentclass[%
 a4paper,11pt
]{article}

\usepackage{amsmath}
\usepackage{graphicx}
\usepackage{dcolumn}% Align table columns on decimal point
\usepackage{bm}% bold math

\usepackage[T1]{fontenc}
\usepackage{jheppub}

\include{macroses}

\title{
Lattice QCD thermodynamics at finite chemical potential
and its comparison with Experiments}

\arxivnumber{1704.03980}

\author[a]{D. Boyda,\note{Corresponding author.}}
\author[a,b]{V.G. Bornyakov}
\author[a]{V. Goy}
\author[a]{A. Molochkov}
\author[a,c]{A. Nakamura}
\author[a]{A. Nikolaev}		
\author[a,d]{V. Zakharov}

\affiliation[a]{School of Biomedicine, Far Eastern Federal University, Russia}
\affiliation[b]{Institute for High Energy Physics NRC Kurchatov Institute, Russia}
\affiliation[c]{Research Center for Nuclear Physics, Osaka University, Japan}
\affiliation[d]{Moscow Institute of Physics and Technology, Dolgoprudny, Russia}

\emailAdd{boyda\_d@mail.ru}

\date{\today}% It is always \today, today,
\abstract{
We compare higher moments of baryon numbers measured at
the RHIC heavy ion collision experiments with 
those by the lattice QCD calculations.
We employ the canonical approach, in which
we can access the real chemical potential
regions avoiding the sign problem.
In the lattice QCD simulations, we study several fits of
the number density in the pure imaginary chemical potential,
and analyze how these fits affects behaviors at
the real chemical potential.
In the energy regions between  $\sqrt{s}_{NN}$=19.6 and 200 GeV,
the susceptibility calculated at $T/T_c=0.93$ is consistent
with experimental data 
at $0 \le \mu_B/T < 1.5$, while
the kurtosis shows similar behavior with that of the experimental data
in the small $\mu_B/T$ regions $0 \le \mu_B/T < 0.3$.
The experimental data at $\sqrt{s}_{NN}=$ 11.5 shows
quite different behavior.
The lattice result in the deconfinement region,$T/T_c=1.35$, is 
far from experimental data.
}

\begin{document}
	\maketitle
	\flushbottom

\section{Introduction}

Many studies have tried to reveal the properties of strongly
interacting quark-gluon/hadron matter from experimental and
phenomenological analyses of high-energy heavy-ion collisions
\cite{adamczyk2014energy,luo2012probing,kharzeev2001hadron,
yokota2016functional}.
It is expected that these studies will lead to understanding
of the phase diagram in the temperature - baryon density plane, which is also looked for in cosmological research.
Although the first principle calculations, based on lattice QCD, 
should provide fundamental information from QCD, 
it had been believed that the ``sign problem'' makes the task
impossible. 

We have been analyzing this ``sign problem'' and have found that
the canonical approach can beat the problem: 
the disease of the canonical approach proposed by Hasenfratz and Toussaint
\cite{hasenfratz1992canonical} can be remedied by the present computing
technique including the multi-precision calculations.
In this paper, we use several computational algorithms 
that we have developed for this purpose. 

We will demonstrate that the canonical approach can
provide data at finite baryon density which can be compared with
those by heavy ion collision data. 
Note that the fugacity expansion, which plays the central role 
in the approach, is not an approximation, but
the exact series on the finite size lattice.

In the next section,
we briefly describe the canonical approach. 
In Section \ref{sec:3}, we show the procedure to compute the canonical partition functions $Z_n$. 
Section \ref{sec:4} is devoted to our results for the physical observables and comparison with respective experimental results. 
Finally, we conclude in Section \ref{sec:5}.

\section{Formulation - Grand canonical partition function 
as a series in terms of canonical partition functions}

The lattice QCD is a simulation study based on the grand canonical partition function,
\beq
Z_{GC}(\mu,T,V) = \int \mathcal{D}U (\det\Delta(\mu))^{N_f} e^{-S_G},
\label{Eq:PathIntegral}
\eeq		
where $\mu$ is the chemical potential, $N_f$ is the number of flavor, 
$S_G$ is a gauge part of action and $\det\Delta$ is the fermion 
determinant, which satisfies the relation
\beq
[\det\Delta(\mu)]^* = \det\Delta(-\mu^*).
\label{Eq:PhaseOfDetDelta}
\eeq
Consequently,
when $\mu$  is nonzero and  real, $\det\Delta$ is complex, and 
when $\mu$ is pure imaginary or zero, $\det\Delta$ is real.

In Monte Carlo simulations, the gluon fields, $U$, are generated
with the probability proportional to the 
integrand
 in Eq.\ (\ref{Eq:PathIntegral}),
and therefore, if $\det\Delta$ is complex, the simulations cannot be conducted.
%If we separate out the phase 
%factor, i.e. rewrite the integrand as
%\beq
%\left( |(\det\Delta)|e^{i\theta} \right)^{N_f}e^{-S_G},
%\label{Eq:SeparationOfPhase}
%\eeq
%and 
%include only the absolute value into the probability,
%then the observables include
%the phase and oscillate.  
This makes the simulation practically impossible,
and is called the ``sign problem''.

In order to circumvent this obstacle, many approaches have been pursued;
see \cite{Karsch2015} for a recent review.
In recent publications \cite{bazavov2017qcd,gunther2016qcd,d2016higher,
datta2016quark} where higher order
cumulants were evaluated for nearly physical quark masses, 
mostly Taylor expansion method was employed and simulations were 
performed 
at zero chemical potential. 
Monte Carlo simulations for pure imaginary $\mu$ are free from
the complex measure problem, as can be seen from Eq.\ (\ref{Eq:PhaseOfDetDelta}). 
The question is {\it how one can extract data for the real $\mu$}.

The grand canonical partition function is related to the canonical
partition function, $Z_C(n,T,V)$, as follows:
\begin{align}
Z_{GC}(\mu,T,V) &= \tr(e^{-\frac{\hat{H} - \mu \hat{N}}{T}})
= \sum_n \bra{n}e^{-\frac{\hat{H}}{T}}\ket{n} e^{\frac{\mu n}{T}}
\nonumber \\
%&=\sum_{n=-\infty}^{\infty} Z_C(n,T,V) e^{\frac{\mu n}{T}}
%=\sum_{n=-\infty}^{\infty} Z_n  e^{\theta n}
= &\sum_{n=-n_{max}}^{n_{max}} Z_n \xi ^n\,,
\label{Eq:ZGC-Zn}
\end{align}
\noindent where
$\xi = e^{\mu/T}$ is the fugacity,
$\hat{N}$ is an operator of a conserved quantum number
such as a baryon number or electric charge.
We introduce an abbreviation $Z_n$ for $Z_C(n,T,V)$. 
Here we use notation $n_{max}$ for number of terms in the sum.
Theoretically, for finite lattice $n_{max} =2 N_c N_x N_y N_z$\cite{nagata2010wilson}, 
but in the real simulations we must truncate the fugacity expansion,
$n<n_{max}$. 
This truncation brings some systematic error. 
In this report, we are mainly concerned with the baryon number case, and we write the chemical potential $\mu_B$.

For imaginary $\mu_B$ ($\mu_B = i \mu_{I}$ and $\theta_I = \mu_{I}/T$),
we can calculate $Z_n$ by the inverse Fourier transformation
~\cite{hasenfratz1992canonical}
as

\begin{eqnarray}
Z_n=\int_0^{2\pi}\frac{d\theta_I}{2\pi}
e^{-in\theta_I}Z_{GC}(\mu_B = i \theta_I T,T,V).
\label{Eq:Zn-Fourier}
\end{eqnarray}

Note that $Z_n= \bra{n} e^{-\frac{\hat{H}}{T}}\ket{n}$ $\ge 0$
does not depend on $\mu_B$, and
therefore one can evaluate the grand canonical partition function,
$Z_{GC}$, in Eq.(\ref{Eq:ZGC-Zn})
for any $\mu_B$ (imaginary or real) once $Z_n$ are known.
After the pioneering work of A.Hasenfraz and Toussant
~\cite{hasenfratz1992canonical}, many approaches in this direction 
were done
~\cite{danzer2012properties,de2006p,li2011critical,alexandru2005lattice}.

The formula Eq.(\ref{Eq:ZGC-Zn}) is exact, and 
it was pointed out in \cite{nagata2010wilson} that
on the finite lattice, $Z_{GC}$ is expressed as
a finite series of the fugacity expansion.

Now we have a route from the imaginary to the real chemical potential
regions:
\begin{itemize}
\item
Step 1: Using Eq.\ (\ref{Eq:Zn-Fourier}),
we calculate $Z_n$ from $Z_{GC}$ 
computed at the imaginary $\mu_B$.
\item
Step 2: Using these $Z_n$ in Eq.\ (\ref{Eq:ZGC-Zn}), we construct $Z_{GC}$
for the real $\mu_B$.
\end{itemize}

To search for the phase transition signals, one can use the moments
$\lambda_{m}$, which can be also extracted from results of the heavy
ion collision experiments:

\beq
\lambda_{m}(\mu_B)=\left(T\frac{\partial}{\partial\mu_B}\right)^m\log Z_{GC}
.
\label{Eq:Moments}
\eeq
Especially, $\lambda_{2}$ (susceptibility), $\lambda_{3}$, and $\lambda_{4}$, provide useful information on the phase structure. In this paper we conduct a detailed study of the canonical ensemble approach which has a potential to reveal the QCD phases. 
We study the systematic error of different natures and its effect to final conclusions. 
We formulate a new approach to determine the range of reliability of the analytical continuation from the imaginary chemical potential to its real values. 
We compute the ratios of the moments $\lambda_m$ in $N_f=2$ lattice QCD and compare results with the values extracted from the RHIC experimental data.

\section{Computation of $Z_n$ \label{sec:3}}

\subsection{Lattice Setup \label{sec:2}}
For simulating the lattice QCD at the imaginary chemical potential, 
we employ the clover improved Wilson fermion action
with two flavors and Iwasaki gauge action.
The details of the simulation were reported in Ref.~\cite{bornyakov2016new}.
Our simulation corresponds to $m_{\pi}/m_{\rho} = 0.8 \quad(m_{\pi} = 0.7$ GeV). 
In this report, we study two temperatures: $T/T_c = 1.35(7)$ ($\beta = 2.0$) corresponds to the deconfinement phase 
and $T/T_c = 0.93(5)$ ($\beta = 1.8$) - the confinement phase.

In the confinement (deconfinement) phase, we make simulations at 27(37) different values of the baryon chemical potential $\mu_{I}$ in the interval $0\le \mu_I/T \le \pi$. Additionally we make simulations at a few values of $\mu_{I}$ above $\pi$ to check the Roberge-Weiss periodicity. 
For each value of $\mu_{I}$, 
1800 configurations separated by ten trajectories are used to evaluate
the physical observables.

\subsection{Direct way to extract $Z_n$ from $n_B$}

Using the standard lattice QCD algorithm,
we can evaluate the baryon number density  $n_B$ directly for any value of
the imaginary chemical potential:
\beq
\frac{n_{B}}{T^{3}}
= i C \int \mathcal{D}U e^{-S_G} (\det\Delta(\mu_{I}))^{N_f}
\tr\left[\Delta^{-1}\frac{\partial \Delta}{\partial \mu_{I}/T}\right],
\label{eq:n_Z}
\eeq
where $C=\frac{N_{f}N_{t}^{3}}{N_s^3 Z_{GC}(0)}$. Note that the number density in imaginary chemical potential regions is pure imaginary.

On the other hand,  the number density is connected with the canonical partition function, $Z_n$ as
\beqa
n_B =  \frac{\lambda_1}{V}
= \frac{T}{V} \frac{\partial}{\partial \mu_B} \, \ln \, Z_{GC}(\mu_B, T)\nonumber\\
= 
\frac{i}{(aN_s)^3}\frac{2  \sum_{n=1}^{n_{max}} Z_n \, n \, \sin(n \theta_{I})}{Z_0 + 2\, \sum_{n=1}^{n_{max}} Z_n \, \cos(n\theta_{I})}\,,
\label{Eq:DefnB1}
\eeqa
\noindent where we used Eq.~(\ref{Eq:ZGC-Zn}) and relation $Z_n = Z_{-n}$.
The direct way to extract the canonical partition functions $Z_n$ from the lattice data for $ n_B $ is to fit the measured baryon number density to Eq. (\ref{Eq:DefnB1}) with $Z_n$ as fitting parameters. 
We tried to do it and realized that the fit goes quite unstable and some $Z_n$'s are negative. 
The difficulty of fitting comes from the drastic cancellations in both the numerator and denominator in Eq.(\ref{Eq:DefnB1}).

\subsection{Construct $Z_{GC}(\theta_I)$ from $n_B$
-- Integration method}

More promising way is to construct the grand canonical partition function from $ n_B $.  
Integrating number density over imaginary $\mu_I$, at fixed temperature $T$, we have
\beqa
\frac{Z_{GC}(\theta_I)}{Z_{GC}(0)}
= \exp \left(V \int_0^{\theta_I}  d (i \tilde {\theta}_{I}) \, i \, \mbox{Im} [n_B (\tilde {\theta}_{I})]
\right)\nonumber\\
= \exp \left(-V \int_0^{\theta_I}  d x \,  n_{BI} (x) \right),
\label{Eq:ZbyIntegrate}
\eeqa
\noindent where we use the fact that $  n_B  $ is pure imaginary and denote $n_{BI} = \mbox{Im}[n_B]$. 
% @Boyda: rewritten with more detail
%We calculate $Z_n$ by inserting this $Z_{GC}$ into Eq~(\ref{Eq:Zn-Fourier}). Then one can construct $Z_{GC}$ as $Z_{GC} = \sum Z_n \xi^n$ at real $\mu_B$. 
According to previously mentioned route we insert above formula into Fourier transformation Eq~(\ref{Eq:Zn-Fourier}) what allows us to calculate canonical functions $Z_n/Z_0$. With these canonical functions we study Grand Canonical Partition Function  $Z_{GC} = \sum Z_n \xi^n$ and other observables (Eq.~\ref{Eq:Moments}) at the real chemical potential.
This procedure provides a new method to study physics in the real chemical potential region via Monte Carlo simulations of the pure imaginary chemical potential \cite{bornyakov2016new, jetpl}.

There is no Ansatz until this point; therefore, Eq.\ (\ref{Eq:ZbyIntegrate}) is exact and theoretically the calculation for any value of the chemical potential is possible. In practice, however, we must introduce some assumptions, and consequently, the reliable range of the real chemical potential values is restricted.

\subsection{Fitting $n_B$ at $T>T_c$ and $T<T_c$}

One way to evaluate the right hand side in Eq.\ (\ref{Eq:ZbyIntegrate}) is to calculate the number density for many values of 
$\mu_I$ and complete the numerical integration. 
In order to obtain a reliable result, we need hundreds of different $\mu_{I}$ values, but this is computationally expensive task. 
In this paper, we employ a 
% @Boyda: simplified -> simple
simplified approach - we fit the numerical data for $ n_B $ and use the fit function in Eq.\ (\ref{Eq:ZbyIntegrate}).
Our task, thus, is how to fit the number density as a function of $\mu_I$:
We employ the Ansatz Eq. (\ref{Eq:nBFourier}) and (\ref{Eq:nBPolinomial}) 
given below, to compute the partition function for imaginary $\mu_B$ from Eq.(\ref{Eq:ZbyIntegrate}).
Then we compute $Z_n$ using Eq. (\ref{Eq:Zn-Fourier}).
This idea is a continuation of another concept in Refs.\ \cite{2004Lombardo,Delia2009,takahashi2015quark} 
- fit the data at imaginary $\mu_B$ and do analytical continuation to the real axis.
The authors of Ref.\ \cite{d2016higher} have recently reported
a thorough analysis.
In Refs.\ \cite{2004Lombardo, Delia2009,takahashi2015quark}, the authors pointed out  that the number density for the imaginary chemical potential is well
approximated by a Fourier series at $T<T_c$,
\beq
n_{BI}(\theta_I)  /T^3 = \sum_{k=1}^{k_{max}} f_{3k} \sin(k\theta_{I}),
\label{Eq:nBFourier}
\eeq
and by a polynomial series at $T>T_c$
\footnote{
At high temperature, there is a Roberge-Weiss phase transition line,
on which $n_{BI}$ becomes singular.
This fact introduces uncertainty for the fitting. 
We discuss this problem
in  the Lee-Yang zero study in near future.
},
\beq
 n_{BI}(\theta_{I}) /T^3 = \sum_{k=1}^{k_{max}} a_{2k-1}\theta_{I}^{2k-1}.
\label{Eq:nBPolinomial}
\eeq
In Refs.~\cite{bornyakov2016new,jetpl}, we confirmed these conclusions with higher precision. 
In computing $Z_{GC}(\theta)/Z_{GC}(0)$, 
we use the parameterizations Eqs. (\ref{Eq:nBFourier}) and (\ref{Eq:nBPolinomial}).
For checking the error, we performed calculations for different $k_{max}$.

To estimate the statistical error, we apply a version of a bootstrap algorithm: here one bootstrap sample consists of a set of standard bootstrap samples of the number density  created for every value of $\mu_{I}$. 
On each bootstrap sample, we fitted data with Eq.~(\ref{Eq:nBFourier}) ($T<T_c$)  or Eq.~(\ref{Eq:nBPolinomial}) ($T>T_c$) and got fitting coefficients $f_{3k}^{i}$ or $a_{2k-1}^{i}$, where $i=1, 2, 3, ..., n_{sample}$ is the sample number. 
Then we calculate $Z_n^{i}$ and physical observable on the bootstrap sample $i$, its average and the error according to the Bootstrap algorithm. Note that, due to fluctuation of the fitting coefficients, the number of positive $Z_n$, i.e.,$n_{max}$, is different on each sample. Details of the modified bootstrap algorithm are described in the Appendix. 
%section \ref{sec:bootstrap}.

Let us comment another approach: the baryon number density fitted to Eqs.\ (\ref{Eq:nBFourier}) or (\ref{Eq:nBPolinomial}) can be analytically continued to the real chemical potential values as was previously studied in Refs.\ \cite{bornyakov2016new,takahashi2015quark,Delia2009,2004Lombardo}.
Baryon number density fit to different functions at imaginary chemical potential with good precision 
can result in substantially different behavior for high enough real $\mu_B$. 
There are no reliable arguments about how to choose unique fitting function.  
For this reason, the range of reliable values of $\mu_B$, where all good fitting functions predict the same results, might be small. 
Consequently, the determination of the reliable range for analytic continuation is problematic.
Another problem of this method is that, the phase transition may be missed because it
implies breaking of analyticity.

On the other hand, the canonical approach provides us with a useful information on the range of reliability at the real chemical potential from the physics. 
From the definition, $Z_n  = \langle n|\exp(-\frac{\hat{H}}{T}) |n\rangle$, 
it follows that $Z_n$ must be positive. 
If $Z_n$ become negative for some $ n > n_{max}$, 
it means that the respective Ansatz for the number density 
used for $Z_n$ calculation via Eq. (\ref{Eq:ZbyIntegrate}) and (\ref{Eq:Zn-Fourier}) can not describe physics for these $n$. 
In the statistical analysis as described above, we used only positive $Z_n$. 
 
 %%%%%%%%%%%%%%%%%%%%%%%%%%%%%%%%%%%%%%%%%%%%%%%%%%%%%%
\begin{figure}[h]
\centering
\includegraphics[scale=0.7]{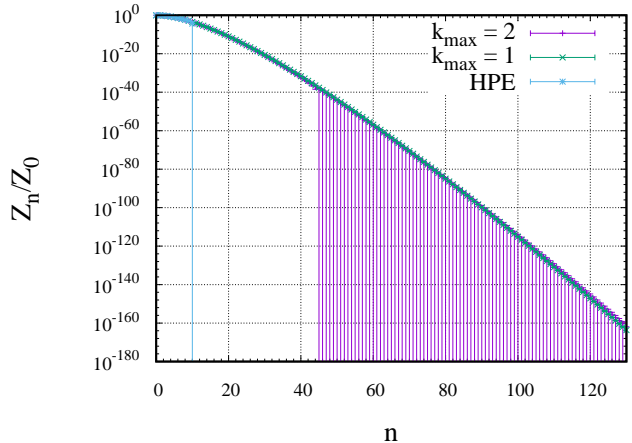}
\caption{Normalized $Z_n$ as a function of $n$ obtained by
the integration method for different number of terms (sines) in Ansatz (Eq. (\ref{Eq:nBFourier})) - $k_{max}$ . 
HPE stands for $Z_n$ calculated using Hopping Parameter Expansion, 
which are taken from \cite{bornyakov2016new}. 
Temperature $T/T_c=0.93$ (Confinement).  
   \label{Fig:AllZnConfinement}   
}
\end{figure}
%%%%%%%%%%%%%%%%%%%%%%%%%%%%%%%%%%%%%%%%%%%%%%%%%%%%%%
  
%%%%%%%%%%%%%%%%%%%%%%%%%%%%%%%%%%%%%%%%%%%%%%%%%%%%%%
\begin{figure}[h]  
\centering
\includegraphics[scale=0.7]{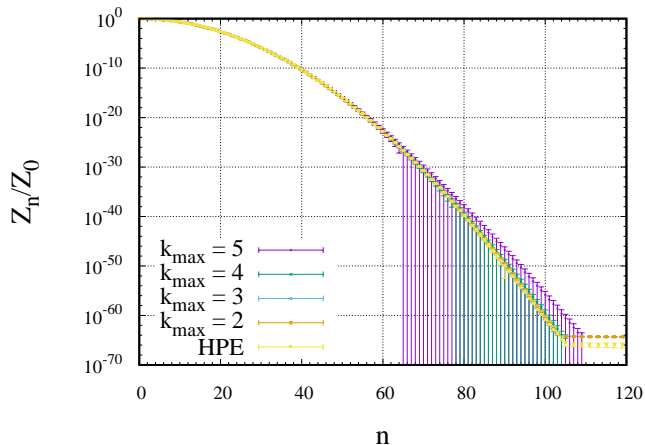}
      \caption{Same as Fig. \ref{Fig:AllZnConfinement} but for temperature $T/T_c=1.35$ (Deconfinement) and polynomial Ansatz (Eq. (\ref{Eq:nBPolinomial})).
  	\label{Fig:AllZnDeConfinement}
}
\end{figure}
%%%%%%%%%%%%%%%%%%%%%%%%%%%%%%%%%%%%%%%%%%%%%%%%%%%%%%

In the Figs.~\ref{Fig:AllZnConfinement} and \ref{Fig:AllZnDeConfinement}, 
one can see $Z_n/Z_0$ calculated with the different number of terms $k_{max}$ 
in Eq.~(\ref{Eq:nBFourier}) and (\ref{Eq:nBPolinomial}). 
At $T/T_c=0.93$ the error of two-term Ansatz is large due to the propagation of large relative error of $f_6$, i.e., noise from our statistics and discrepancy between $k_{max}=1$ and $k_{max}=2$ is small. 
At $T/T_c=1.35$ the behavior for varying $k_{max}$ is similar. For comparison we also present $Z_n$ calculated with winding number expansion method in hopping parameters expansion approximation (labeled by HPE on the figures) borrowed from Ref. \cite{bornyakov2016new}.
For the polynomial fits starting from some $n=n_{max}$ value, the sign of $Z_n$ starts to alternate, while the absolute value changes very slowly. 
Thus we conclude that polynomial series can describe the data only for relatively small fixed $n_{max}$ or $\mu_{reliable}$.
For the fits (\ref{Eq:nBFourier}) with even $k_{max}$ similar thing happens: starting from some $n_{max}$ there appears alternation of the sign of $Z_n$.  
We thus conclude that the Fourier fits with even $k_{max}$ should be also avoided or used for the range of $n$ values below respective $n_{max}$.

\section{Baryon number fluctuations 
-- Comparison of lattice QCD with experimental data \label{sec:4}}

\subsection{Baryon number density}
%%%%%%%%%%%%%%%%%%%%%%%%%%%%%%%%%%%%%%%%%%%%%%%%%%%%%%
\begin{figure}[h]
  \centering
       \includegraphics[scale=0.7]{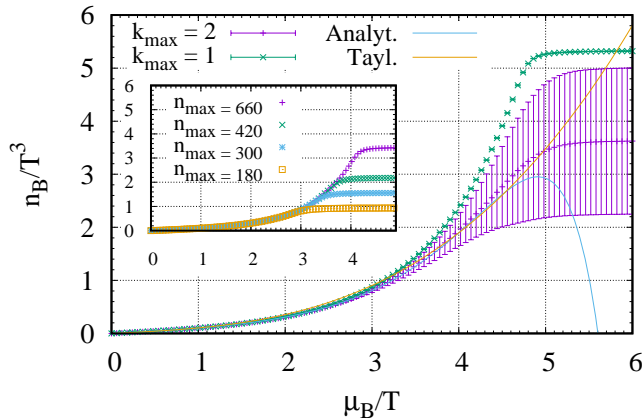}
        \caption{The baryon number density for the real
	chemical potential at $T/T_c=0.93$ calculated using integration method with different number of terms $k_{max}$ in Ansatz Eq. (\ref{Eq:nBFourier}). At insert the effect of finite number of $Z_n$ - $n_{max}$ is shown for $k_{max}=2$. For comparison we also present direct analytical continuation (for $k_{max}=2$) and Taylor expansion method with two coefficients $c_1$ and $c_2$ taken from \cite{WHOT2010}.
        \label{Fig:ReBarDens_Nt4Ns16T0.93Tc} % Fig.5
        }
\end{figure}  
\begin{figure}[h]      
  \centering
        \includegraphics[scale=0.7]{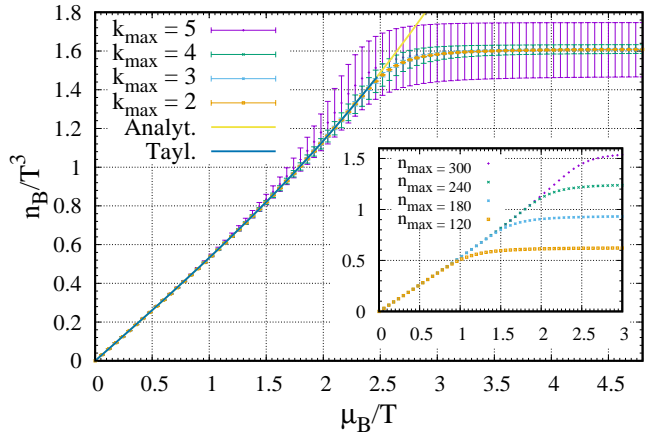}
        \caption{Same as Fig. \ref{Fig:ReBarDens_Nt4Ns16T0.93Tc} but for temperature $T/T_c=1.35$ and Ansatz Eq.(\ref{Eq:nBPolinomial})
        \label{Fig:ReBarDens_Nt4Ns16T1_35Tc}
        }
\end{figure}
%%%%%%%%%%%%%%%%%%%%%%%%%%%%%%%%%%%%%%%%%%%%%%%%%%%%%%

The canonical partition functions, $Z_n$, are now obtained 
using the Fourier Ansatz in the confinement phase and 
the polynomial Ansatz in the deconfinement phase with different $k_{max}$. 
Now, we calculate the baryon number density for the real baryon chemical potential $\mu_B$ values via Eq.~(\ref{Eq:DefnB1}) 
%@Boyda
%	with restricted number of terms to $n_{max}$.
%	->
with fixed number of physical $Z_n$ and $n_{max}$.
The results are shown in Figs.~\ref{Fig:ReBarDens_Nt4Ns16T0.93Tc} and \ref{Fig:ReBarDens_Nt4Ns16T1_35Tc}. 
For large $\mu_B/T$, the number density obtained via Eq.~(\ref{Eq:DefnB1}) goes to plateau as a consequence of the finite number terms in respective sums. 
This effect allows us to 
% @Boyda
%	estimate reliability range of our calculation in real $\mu_B$. 
%	->
study reliability range at real region as a function of $n_{max}$.
The position $\mu_{reliable}/T$ where baryon density goes to plateau is determined by $n_{max}$: the bigger $n_{max}$ is, the bigger $\mu_{reliable}/T$. 
On the insets of Figs. 
\ref{Fig:ReBarDens_Nt4Ns16T0.93Tc} 
and 
\ref{Fig:ReBarDens_Nt4Ns16T1_35Tc}, 
we present data with different $n_{max}$: results coincide up to $\mu_{reliable}/T$ where number density goes to constant. 
Therefore, we conclude that systematical error due to the truncation 
in (\ref{Eq:ZGC-Zn}) affects only on the determination of $\mu_{reliable}/T$,
and do not change results in the reliable range.
We expect that increasing of statistics will allow us to increase $n_{max}$.

From Figs.  \ref{Fig:ReBarDens_Nt4Ns16T0.93Tc} and \ref{Fig:ReBarDens_Nt4Ns16T1_35Tc}, 
it is clear that our data for baryon density becomes unreliable 
after $\mu_{reliable}/T \sim 4-5$ at $T/T_c=0.93$ and $\mu_{reliable}/T \sim 2-2.5$ 
at $T/T_c=1.35$. 
For comparison, we also show the data for direct analytical continuation ($k_{max}$=2) 
and Taylor expansion method. Taylor coefficients were borrowed form \cite{WHOT2010}.

%%%%%%%%%%%%%%%%%%%%%%%%%%%%%%%%%%%%%%%%%%%%%%%%%%%%%%
\begin{figure}[ht]
  \centering
        \includegraphics[scale=0.7]{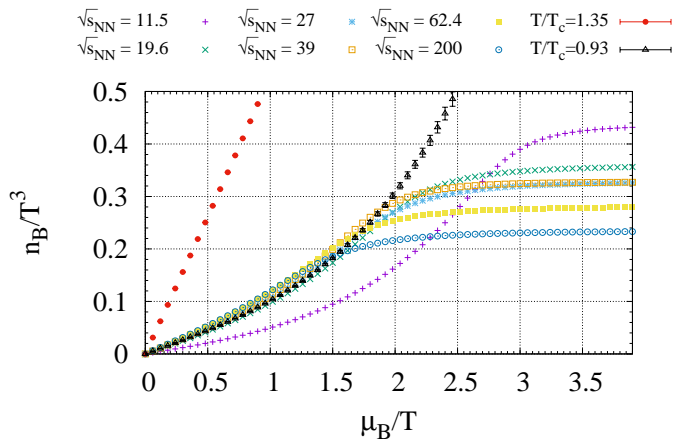}
        \caption{
	Baryon number density for RHIC data of different energies $\sqrt{s_{NN}}$ in GeV together with our lattice results for temperatures $T/T_c=0.93$ and $T/T_c=1.35$.
  	\label{Fig:nBRHIC}
	}

\end{figure}
%%%%%%%%%%%%%%%%%%%%%%%%%%%%%%%%%%%%%%%%%%%%%%%%%%%%%%

The canonical partition functions, $Z_n$, can be directly extracted from an experiment on heavy ions collision. 
Indeed, the multiplicity distributions, $P_n$, measured in the experiment 
have a meaning of the probability and consequently connected with $Z_n$: $P_n=Z_n \xi^n$. 
Following this argument in Ref.\ \cite{nakamura2016probing}, 
$Z_n$ as well as fugacity were extracted from the RHIC experiments data. 
The authors compared extracted fugacity with different works on estimation of freeze-out parameters and got agreement within 5 \%.
With $Z_n$ extracted from RHIC data \cite{nakamura2016probing}, we construct the baryon number densities by Eq.(\ref{Eq:DefnB1}) for each RHIC energy. 
See Fig.\ref{Fig:nBRHIC}, where we show also our lattice results at $T/T_c=$0.93 (black) and 1.35 (red). 
We observe
\begin{itemize}
\item
the baryon number density calculated on the lattice at $T>T_c$ deviates from RHIC data, 
\item
in the confinement, the experimental data are consistent with the lattice calculation except $\sqrt{s_{NN}}$=11.5 GeV. 
\end{itemize}
Note that
the estimated temperature at $\sqrt{s_{NN}}$=11.5 GeV in Ref.\cite{Alba:2014eba} is significantly below other energy data. 

\subsection{Moments of the baryon number density}

%%%%%%%%%%%%%%%%%%%%%%%%%%%%%%%%%%%%%%%%%%%%%%%%%%%%%%
\begin{figure}[h]
  \centering
        \includegraphics[scale=0.7]{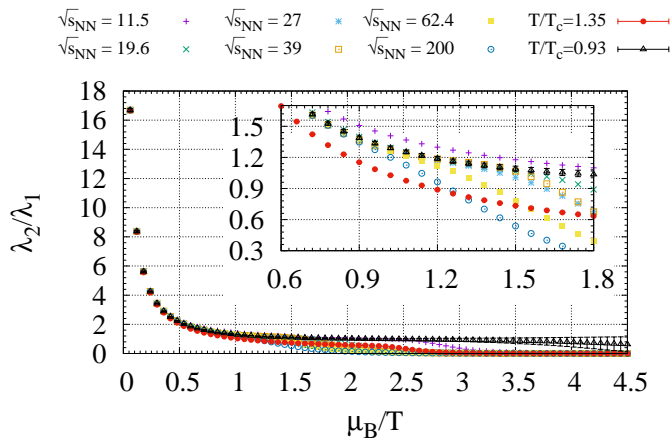}
        \caption{
	Same as Fig. \ref{Fig:nBRHIC} but for $\lambda_2/\lambda_1$.
The inset figure shows the region $0.6 \le \mu_B/T \le 1.8$
to see recognizably the behavior of the lattice QCD result at $T/T_c=0.93$ (black)
and the RHIC energy scan data.
\label{Fig:R21T0_93Tc}
	}
\end{figure}
%%%%%%%%%%%%%%%%%%%%%%%%%%%%%%%%%%%%%%%%%%%%%%%%%%%%%%

\begin{figure}[h]
  \centering
        \includegraphics[scale=0.7]{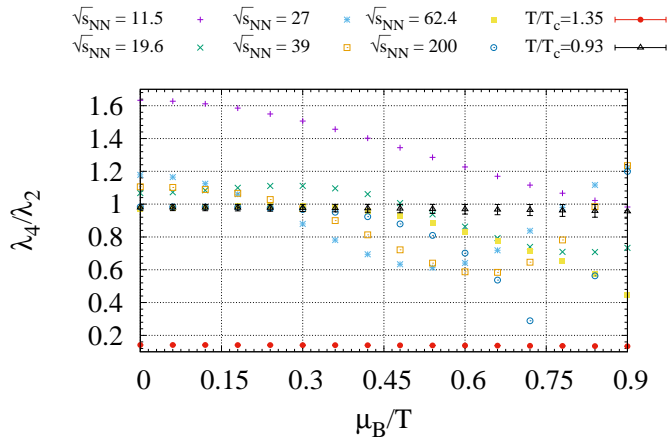}
        \caption{
		Same as Fig. \ref{Fig:nBRHIC} but for 	$\lambda_4/\lambda_2$.
		The ratio, $\lambda_4/\lambda_2$, at $T=1.35T_c$ (red) decreases slowly as 
		$\mu_B/T$ increases.
		}
        \label{Fig:R42T0_93Tc}
\end{figure}

Observables $\lambda_2/\lambda_1$ and $\lambda_4/\lambda_2$ can be constructed from calculated $Z_n$. 
In Figs.\ \ref{Fig:R21T0_93Tc} and \ref{Fig:R42T0_93Tc}, we show these ratios as calculated by the integration method described above together with those extracted from the RHIC Star data.

In the relativistic heavy-ion collision experiments, $\lambda_2/\lambda_1$ (susceptibility) and $\lambda_4/\lambda_2$ (kurtosis) are expected to be good indicators for detecting the QCD phase transition \cite{luo2012probing,redlich2012probing}. We calculate these quantities and compare with experiments. 
% @Boyda: nex paragragh rewritten
Note that $Z_n$ were constructed from the proton multiplicity data, not the baryon multiplicity and our data were extracted for high pion mass. Thus  one should consider our results as a proxy for the real baryon number moments. \textit{Nevertheless our data, presented in Figs. 7 and 8,  agreed with state of the art Taylor expansion results for physical quark masses \cite{2017Bazarov} as it was shown in \cite{2017BornyakovEPJ}.}

% @Boyda
In Fig.~\ref{Fig:R21T0_93Tc}, the ratio $\lambda_2/\lambda_1$ (susceptibility) 
at small $\mu/T$ decreases rapidly; Then it behaves as a constant and finally it drops to zero as $\mu/T$ becomes large. 
The rapid drop from one to zero in $\lambda_2/\lambda_1$ may indicate the phase transition, but this is not the case. 
This is because the dropping position shifts with varying of $n_{max}$. 
This tells us $\mu_{reliable}$ for this observable.
Another observable $\lambda_4/\lambda_2$ (kurtosis) as well as $\lambda_2/\lambda_1$ can indicate phase transition. As shown on Fig.~\ref{Fig:R42T0_93Tc} it has constant value at small $\mu/T$ and decreases then. No indication of phase transition can be seen.

%%%%%%%%%%%%%%%%%%%%%%%%%%%%%%%%%%%%%%%%%%%%%%%%%%%%%%
%%%%%%%%%%%%%%%%%%%%%%%%%%%%%%%%%%%%%%%%%%%%%%%%%%%%%%

In the confinement regions, we see very good agreement for $\lambda_2/\lambda_1$ and $\lambda_4/\lambda_2$ between the lattice calculation and those estimated from RHIC data.
In the energy regions between  $\sqrt{s}_{NN}$=19.6 and 200 GeV, the susceptibility calculated at $T/T_c=0.93$ is consistent with experimental data  at $0 \le \mu_B/T < 1.5$, while the kurtosis shows similar behavior with that of the experimental data in the small $\mu_B/T$ regions $0 \le \mu_B/T < 0.3$. The experimental data at $\sqrt{s}_{NN}=$ 11.5 shows
quite different behavior. The lattice data in the deconfinement region, $T/T_c=1.35$, is 
far from experimental data. 
Note that for RHIC data, one has only small number of $Z_n$ (i.e., $n_{max}$);
Thus reliability range for experiment at Figs.~\ref{Fig:R21T0_93Tc} and~\ref{Fig:R42T0_93Tc} is %much smaller then for our data.
very limited.

\section{Concluding Remarks \label{sec:5}}

In this paper, we study an approach for revealing the QCD phase structure using lattice QCD simulations. 
Prior to this study, it was believed that this was impossible because of the sign problem; only small density regions could be studied by extrapolating from the data at $\mu_B/T=0$.
However, all relevant information on the QCD phase at finite baryon density is contained in the imaginary chemical potential regions, $0 \le \mu_{I}/T \le \pi$. 
The question is how to map this information to the real chemical potential. Eq.\ (\ref{Eq:ZGC-Zn}) provides a possible solution, 
because $Z_n$ can be calculated in the imaginary chemical potential regions. 
Since numerical Monte Carlo simulations provide results with finite accuracy, 
we should find practical methods which work.

%We fit the number density at the imaginary chemical potential using $Z_n$ as parameters and found that it does not work.

We parametrize the number density in the imaginary chemical potential with the Ansatz 
(the Fourier series in the confinement and polynomial series in the deconfinement) 
and integrate them to get the grand partition function.  
The canonical partition function, $Z_n$, and other observables are then calculated from them.

This method produces $Z_n$ up to $n_{max}$ which is determined by the used Ansatz and current statistics. However, this is not the first principle calculation because of introducing an assumption to the number density. 
Analyzing $n_{max}$ behavior we can estimate reliability of this assumption. 
On the other hand, better interpolation procedure (cubic spline for example) or numerical integration rather than any Ansatz release our data from assumptions. 
This study will be reported in future.

%%%%% Added
In the paper, we studied how adding terms affect results at real $\mu$. 
From Fig. \ref{Fig:ReBarDens_Nt4Ns16T0.93Tc}, we observe that 
(1) two {\it sine} and one {\it sine} results agreed in two sigma, but it seems there is small signal; 
(2) adding the second term even with small coefficients change results 
and affect in drastic error rising. 

Recently, similar approaches to the QCD phase transition were investigated
\cite{PhysRevD.97.114030},
\cite{Almasi:2018lok}.
In \cite{PhysRevD.97.114030}, the coefficients of the fugacity expansion 
of the pressure are parametrized, while
in \cite{Almasi:2018lok}, the Fourier coefficients of the baryon number
density are parametrized.
These analyses may open the way to study the criticalities of the QCD at
finite temperature and baryon density.

We studied how errors of calculations made at the imaginary chemical potential propagate into errors at the real one. 
We also found range of reliability of our results by imposing the condition that $Z_n$ have to be positive.

We then investigate whether we can estimate $\lambda_2/\lambda_1$ and $\lambda_4/\lambda_2$. 
The results are consistent with the values estimated from the RHIC experiments as shown in Figs.\ \ref{Fig:R21T0_93Tc} and \ref{Fig:R42T0_93Tc}. 
This is very encouraging.
More realistic simulations with the physical quark mass and small lattice spacing, may even predict the temperature of the experimental data.

When we map the information from the pure imaginary to the real chemical potential by the canonical method, the reliable regions for baryon density is $\mu_{reliable}/T \sim 4-5$ at $T/T_c=0.93$ and $\mu_{reliable}/T \sim 2-2.5$ at $T/T_c=0.93$.  This limitation comes from the finite numbers of $Z_n$ as a consequence of polynomial (in the deconfinement phase) or Fourier (in the confinement phase) Ansatz. They can be increased by increasing of statistics.

\bigskip
\section*{Acknowledgment}
This work was completed thanks to support from RSF grant 15-12-20008.
Work done by A. Nakamura on the theoretical formulation of $Z_n$
for comparison with experiments was supported by JSPS KAKENHI
Grant Numbers 26610072 and 15H03663.
The calculations were performed on Vostok-1 at FEFU.

\section*{Appendix - Modified Bootstrap algorithm. \label{sec:bootstrap}}

We describe modified bootstrap algorithm by which the errors are estimated in this paper. As raw statistical data we have Monte Carlo samples of number density $n^i_{MC}(\mu_k)$, where $i=1,...,N_{MC}$ and $\mu_k \in [-\pi/3;\pi/3]$. Our final observables are canonical partition functions $Z_n$ and different thermodynamical quantities. Due to our method described above final observables depend on raw MC data non-locally ($Z_n$ calculation requires number density for different $\mu_k$) thus we had to implement modified Bootstrap algorithm. We describe our algorithm as following:
\begin{enumerate}
	\item Fix $\mu_k$ and recreate bootstrap samples $n^t_{BS}(\mu_k)$ ($t = 1, 2, ..., N_{bs}$) from raw MC configuration $n^i_{MC}(\mu_k)$ using standard algorithm \cite{Gattringer}: randomly choose $N_{MC}$ configuration and calculate average.
	\item Repeat the previous step for all values of $\mu_k$  in the range $[-\pi/3;\pi/3]$ which gives $N_{bs}$ Bootstrap samples for all set of $\mu_k$.
	\item For fixed $t$ (Bootstrap sample number) and different $\mu_k$ fit data $n^t_{BS}(\mu_k)$ as a function of $\mu$. As a result one has $N_{fit}$ coefficients of fitting function $f_p^t$, where $p=1,2,...,N_{fit}$.
	\item Using $N_{fit}$ fitting coefficients $f_p^t$ with fixed $t$ calculate $Z^t_n$ with Integral method.
	\item From $Z^t_n$ calculate thermodynamical observables if any.
	\item Repeat steps 3 - 5 for $t=1,2,... N_{BS}$ and calculate error with Bootstrap algorithm rules: $$ \delta O = \sqrt{ \frac{1}{N_{BS}}\sum_t (O_t - \bar{O})^2 },$$ where $O$ is $Z_n$ or any thermodynamical quantity, $\bar{O}$ is average of $O$.
\end{enumerate}

\providecommand{\href}[2]{#2}\begingroup\raggedright\endgroup

\end{document}

%% file: macroses.tex
\newcommand{\tr}{ {\rm Tr} \, }

\newcommand{\ket}[1]{ \, | #1 \rangle }
\newcommand{\bra}[1]{ \langle #1 | \, }

\newcommand{\beq}{\begin{equation}}
\newcommand{\eeq}{\end{equation}}
\newcommand{\beqa}{\begin{eqnarray}}
\newcommand{\eeqa}{\end{eqnarray}}

%% file: manuscript_JHEP.bbl
\begin{thebibliography}{10}
	
	\bibitem{adamczyk2014energy}
	{\scshape STAR Collaboration} collaboration, L.~Adamczyk, J.~K. Adkins,
	G.~Agakishiev, M.~M. Aggarwal, Z.~Ahammed, I.~Alekseev et~al., \emph{{Energy
			Dependence of Moments of Net-Proton Multiplicity Distributions at RHIC}},
	\href{https://doi.org/10.1103/PhysRevLett.112.032302}{\emph{Phys. Rev. Lett.}
		{\bfseries 112} (2014) 032302}.
	
	\bibitem{luo2012probing}
	X.~Luo, \emph{{Probing the QCD critical point by higher moments of net-proton
			multiplicity distributions at STAR}}, {\emph{Open Physics} {\bfseries 10}
		(2012) 1372}.
	
	\bibitem{kharzeev2001hadron}
	D.~Kharzeev and M.~Nardi, \emph{{Hadron production in nuclear collisions at
			RHIC and high-density QCD}}, {\emph{Physics Letters B} {\bfseries 507} (2001)
		121}.
	
	\bibitem{yokota2016functional}
	T.~Yokota, T.~Kunihiro and K.~Morita, \emph{{Functional renormalization group
			analysis of the soft mode at the QCD critical point}}, {\emph{Progress of
			Theoretical and Experimental Physics} {\bfseries 2016} (2016) 073D01}.
	
	\bibitem{hasenfratz1992canonical}
	A.~Hasenfratz and D.~Toussaint, \emph{{Canonical ensembles and nonzero density
			quantum chromodynamics}}, {\emph{Nuclear Physics B} {\bfseries 371} (1992)
		539}.
	
	\bibitem{Karsch2015}
	H.-T. {Ding}, F.~{Karsch} and S.~{Mukherjee}, \emph{{{Thermodynamics of
				strong-interaction matter from lattice QCD}}},
	\href{https://doi.org/10.1142/S0218301315300076}{\emph{International Journal
			of Modern Physics E} {\bfseries 24} (2015) 1530007}
	[\href{https://arxiv.org/abs/1504.05274}{{\ttfamily 1504.05274}}].
	
	\bibitem{bazavov2017qcd}
	A.~Bazavov, H.-T. Ding, P.~Hegde, O.~Kaczmarek, F.~Karsch, E.~Laermann et~al.,
	\emph{{QCD equation of state to $\mathcal{O}({\ensuremath{\mu}}_{B}^{6})$
			from lattice QCD}},
	\href{https://doi.org/10.1103/PhysRevD.95.054504}{\emph{Phys. Rev. D}
		{\bfseries 95} (2017) 054504}.
	
	\bibitem{gunther2016qcd}
	J.~Gunther, R.~Bellwied, S.~Borsanyi, Z.~Fodor, S.~Katz, A.~Pasztor et~al.,
	\emph{{The QCD equation of state at finite density from analytical
			continuation}}, {\emph{arXiv preprint arXiv:1607.02493} (2016) }.
	
	\bibitem{d2016higher}
	M.~D'Elia, G.~Gagliardi and F.~Sanfilippo, \emph{{Higher order quark number
			fluctuations via imaginary chemical potentials in $ N_f= 2+ 1$ QCD}},
	{\emph{arXiv preprint arXiv:1611.08285} (2016) }.
	
	\bibitem{datta2016quark}
	S.~Datta, R.~V. Gavai and S.~Gupta, \emph{{Quark number susceptibilities and
			equation of state at finite chemical potential in staggered QCD with $N_t=
			8$}}, {\emph{arXiv preprint arXiv:1612.06673} (2016) }.
	
	\bibitem{nagata2010wilson}
	K.~Nagata and A.~Nakamura, \emph{Wilson fermion determinant in lattice qcd},
	{\emph{Physical Review D} {\bfseries 82} (2010) 094027}
	[\href{https://arxiv.org/abs/1009.2149}{{\ttfamily 1009.2149}}].
	
	\bibitem{danzer2012properties}
	J.~Danzer and C.~Gattringer, \emph{Properties of canonical determinants and a
		test of fugacity expansion for finite density lattice qcd with wilson
		fermions}, {\emph{Physical Review D} {\bfseries 86} (2012) 014502}
	[\href{https://arxiv.org/abs/1204.1020}{{\ttfamily 1204.1020}}].
	
	\bibitem{de2006p}
	P.~de~Forcrand, \emph{P. de forcrand and s. kratochvila, nucl. phys. b, proc.
		suppl. 153, 62 (2006).}, {\emph{Nucl. Phys. B, Proc. Suppl.} {\bfseries 153}
		(2006) 62} [\href{https://arxiv.org/abs/0602024}{{\ttfamily 0602024}}].
	
	\bibitem{li2011critical}
	A.~Li, A.~Alexandru, K.-F. Liu et~al., \emph{Critical point of n f= 3 qcd from
		lattice simulations in the canonical ensemble}, {\emph{Physical Review D}
		{\bfseries 84} (2011) 071503}
	[\href{https://arxiv.org/abs/1103.3045}{{\ttfamily 1103.3045}}].
	
	\bibitem{alexandru2005lattice}
	A.~Alexandru, M.~Faber, I.~Horv{\'a}th and K.-F. Liu, \emph{Lattice qcd at
		finite density via a new canonical approach}, {\emph{Physical Review D}
		{\bfseries 72} (2005) 114513}
	[\href{https://arxiv.org/abs/0507020}{{\ttfamily 0507020}}].
	
	\bibitem{bornyakov2016new}
	V.~Bornyakov, D.~Boyda, V.~Goy, A.~Molochkov, A.~Nakamura, A.~Nikolaev et~al.,
	\emph{New approach to canonical partition functions computation in $ n_f= 2$
		lattice qcd at finite baryon density}, {\emph{Phys. Rev. D} {\bfseries 95}
		(2017) 094506} [\href{https://arxiv.org/abs/1611.04229}{{\ttfamily
			1611.04229}}].
	
	\bibitem{jetpl}
	D.~L. Boyda, V.~G. Bornyakov, V.~A. Goy, V.~I. Zakharov, A.~V. Molochkov,
	A.~Nakamura et~al., \emph{Novel approach to deriving the canonical generating
		functional in lattice qcd at a finite chemical potential},
	\href{https://doi.org/10.1134/S0021364016220069}{\emph{JETP Letters}
		{\bfseries 104} (2016) 657}.
	
	\bibitem{2004Lombardo}
	M.~D'Elia and M.-P. Lombardo, \emph{Qcd thermodynamics from an imaginary
		${\ensuremath{\mu}}_{B}$: Results on the four flavor lattice model},
	\href{https://doi.org/10.1103/PhysRevD.70.074509}{\emph{Phys. Rev. D}
		{\bfseries 70} (2004) 074509}
	[\href{https://arxiv.org/abs/0406012}{{\ttfamily 0406012}}].
	
	\bibitem{Delia2009}
	M.~{D'Elia} and F.~{Sanfilippo}, \emph{{{Thermodynamics of two flavor QCD from
				imaginary chemical potentials}}},
	\href{https://doi.org/10.1103/PhysRevD.80.014502}{\emph{Physical Review D}
		{\bfseries 80} (2009) 014502}
	[\href{https://arxiv.org/abs/0904.1400}{{\ttfamily 0904.1400}}].
	
	\bibitem{takahashi2015quark}
	J.~Takahashi, H.~Kouno and M.~Yahiro, \emph{{Quark number densities at
			imaginary chemical potential in $N_f = 2$ lattice QCD with Wilson fermions
			and its model analyses}}, {\emph{Physical Review D} {\bfseries 91} (2015)
		014501}.
	
	\bibitem{WHOT2010}
	{\scshape WHOT-QCD} collaboration, S.~Ejiri, Y.~Maezawa, N.~Ukita, S.~Aoki,
	T.~Hatsuda, N.~Ishii et~al., \emph{{{Equation of State and Heavy-Quark Free
				Energy at Finite Temperature and Density in Two Flavor Lattice QCD with
				Wilson Quark Action}}},
	\href{https://doi.org/10.1103/PhysRevD.82.014508}{\emph{Phys. Rev.}
		{\bfseries D82} (2010) 014508}
	[\href{https://arxiv.org/abs/0909.2121}{{\ttfamily 0909.2121}}].
	
	\bibitem{nakamura2016probing}
	A.~Nakamura and K.~Nagata, \emph{{Probing QCD phase structure using baryon
			multiplicity distribution}}, {\emph{Progress of Theoretical and Experimental
			Physics} {\bfseries 2016} (2016) 033D01}.
	
	\bibitem{Alba:2014eba}
	P.~Alba, W.~Alberico, R.~Bellwied, M.~Bluhm, V.~Mantovani~Sarti, M.~Nahrgang
	et~al., \emph{{{Freeze-out conditions from net-proton and net-charge
				fluctuations at RHIC}}},
	\href{https://doi.org/10.1016/j.physletb.2014.09.052}{\emph{Phys. Lett.}
		{\bfseries B738} (2014) 305}
	[\href{https://arxiv.org/abs/1403.4903}{{\ttfamily 1403.4903}}].
	
	\bibitem{redlich2012probing}
	K.~Redlich, \emph{{Probing the QCD chiral cross-over transition in heavy ion
			collisions}}, \href{https://doi.org/10.2478/s11534-012-0105-0}{\emph{Central
			European Journal of Physics} {\bfseries 10} (2012) 1254}.
	
	\bibitem{2017Bazarov}
	A.~Bazavov, H.-T. Ding, P.~Hegde, O.~Kaczmarek, F.~Karsch, E.~Laermann et~al.,
	\emph{Qcd equation of state to $\mathcal{O}({\ensuremath{\mu}}_{B}^{6})$ from
		lattice qcd}, \href{https://doi.org/10.1103/PhysRevD.95.054504}{\emph{Phys.
			Rev. D} {\bfseries 95} (2017) 054504}.
	
	\bibitem{2017BornyakovEPJ}
	{Bornyakov, V.G.}, {Boyda, D.L.}, {Goy, V.A.}, {Iida, H.}, {Molochkov, A.V.},
	{Nakamura, Atsushi} et~al., \emph{Lattice qcd at finite baryon density using
		analytic continuation},
	\href{https://doi.org/10.1051/epjconf/201818202017}{\emph{EPJ Web Conf.}
		{\bfseries 182} (2018) 02017}.
	
	\bibitem{PhysRevD.97.114030}
	V.~Vovchenko, J.~Steinheimer, O.~Philipsen and H.~Stoecker, \emph{Cluster
		expansion model for qcd baryon number fluctuations: No phase transition at
		${\ensuremath{\mu}}_{B}/t<\ensuremath{\pi}$},
	\href{https://doi.org/10.1103/PhysRevD.97.114030}{\emph{Phys. Rev. D}
		{\bfseries 97} (2018) 114030}
	[\href{https://arxiv.org/abs/1711.01261}{{\ttfamily 1711.01261}}].
	
	\bibitem{Almasi:2018lok}
	G.~A. Almasi, B.~Friman, K.~Morita, P.~M. Lo and K.~Redlich, \emph{{Fourier
			coefficients of the net-baryon number density and chiral criticality}},
	\href{https://arxiv.org/abs/1805.04441}{{\ttfamily 1805.04441}}.
	
	\bibitem{Gattringer}
	C.~Gattringer and C.~B. Lang, \emph{{Quantum chromodynamics on the lattice}},
	Lecture Notes in Physics. Springer, Berlin Heidelberg, 2010.
	
\end{thebibliography}
